\documentclass[runningheads,a4paper]{llncs}

\usepackage{caption}

\usepackage{amssymb}
\usepackage{graphicx}
\usepackage{verbatim}
\usepackage{amsmath}
\usepackage{url}
\usepackage{cite}
\usepackage{listings}
\usepackage{algorithm}
\usepackage{algpseudocode}
\usepackage{placeins}
\usepackage{flafter}
\usepackage{tabularx}
\usepackage{subfigure}
\usepackage{txfonts}

\usepackage{rotating}
\urldef{\mailsa}\path|{alfred.hofmann, ursula.barth, ingrid.haas, frank.holzwarth,|
\urldef{\mailsb}\path|anna.kramer, leonie.kunz, christine.reiss, nicole.sator,|
\urldef{\mailsc}\path|erika.siebert-cole, peter.strasser, lncs}@springer.com|    
\newcommand{\keywords}[1]{\par\addvspace\baselineskip

\newcommand{\qed}{\nobreak \ifvmode \relax \else
      \ifdim\lastskip<1.5em \hskip-\lastskip
      \hskip1.5em plus0em minus0.5em \fi \nobreak
      \vrule height0.75em width0.5em depth0.25em\fi}
      
\noindent\keywordname\enspace\ignorespaces#1}

\newcommand{\hide}[1]{}

\title{Formal Modeling and Analysis of Pancreatic Cancer Microenvironment}

\begin{document}

\mainmatter  

\vspace{-.2cm}
\author{Qinsi Wang$^1$, Natasa Miskov-Zivanov$^2$, Bing Liu$^3$, \\James R. Faeder$^3$, Michael Lotze$^4$, Edmund M. Clarke$^1$}

\authorrunning{Q. Wang et al.}
\institute{$^1$ Computer Science Department, Carnegie Mellon University, USA\\
$^2$ Electrical and Computer Engineering Department, Carnegie Mellon University, USA\\
$^3$ Department of Computational and Systems Biology, University of Pittsburgh, USA\\
$^4$ Surgery and Bioengineering, UPMC, USA}
\maketitle
\vspace{-.5cm}
\begin{abstract}
The focus of pancreatic cancer research has been shifted from pancreatic cancer cells towards their microenvironment, involving pancreatic stellate cells that interact with cancer cells and influence tumor progression. To quantitatively understand the pancreatic cancer microenvironment, we construct a computational model for intracellular signaling networks of cancer cells and stellate cells as well as their intercellular communication. We extend the rule-based BioNetGen language to depict intra- and inter-cellular dynamics using discrete and continuous variables respectively. Our framework also enables a statistical model checking procedure for analyzing the system behavior in response to various perturbations. The results demonstrate the predictive power of our model by identifying important system properties that are consistent with existing experimental observations. We also obtain interesting insights into the development of novel therapeutic strategies for pancreatic cancer.   
\end{abstract}
\vspace{-.9cm}
\section{Introduction} 
\vspace{-.2cm}
Pancreatic cancer (PC), as an extremely aggressive disease, is the seventh leading cause of cancer death globally \cite{who2014}. For decades, extensive efforts were made on developing therapeutic strategies targeting at pancreatic cancer cells (PCCs). However, the poor prognosis for PC remains largely unchanged. Recent studies have revealed that the failure of systemic therapies for PC is partially due to the tumor microenvironment, which turns out to be essential to PC development  \cite{kleeff2007pancreatic, erkan2010tumor, farrow2008role, feig2012pancreas}. As a characteristic feature of PC, the microenvironment includes pancreatic stellate cells (PSCs), immune cells, endothelial cells, nerve cells, lymphocytes, dendritic cells, the extracellular matrix, and other molecules surrounding PCCs, among which, PSCs play key roles during the PC development \cite{kleeff2007pancreatic}. In this paper, to obtain a system-level understanding of the PC microenvironment, we construct a multicellular model including intracellular signaling networks of PCCs and PSCs respectively, and intercellular interactions among them.

Boolean Networks (BNs) \cite{thomas1995dynamical} has been widely used to model biological networks \cite{albert2014boolean}. A Boolean network is an executable model that characterizes the status of each biomolecule by a binary variable that related to the abundance or activity of the molecule. It can capture the overall behavior of a biological network and provide important insights and predictions. Recently, it has been found useful to study the signaling networks in PCCs \cite{gong2011symbolic, gong2011formal}. Rule-based modeling language is another successfully used formalism for dynamical biological systems, which allows molecular/kinetic details of signaling cascades to be specified \cite{faeder2009rule, danos2007rule}. It provides a rich yet concise description of signaling proteins and their interactions by representing interacting molecules as structured objects and by using pattern-based rules to encode their interactions. The dynamics of the underlying system can be tracked by performing stochastic simulations. In this paper, to formally describe our multicellular and multiscale model, we extend the rule-based language BioNetGen \cite{faeder2009rule} to enable the formal specification of not only the signaling network within a single cell, but also interactions among multiple cells. Specifically, we represent the intercellular level dynamics using rules with continuous variables and use BNs to capture the dynamics of intracellular signaling networks, considering the fact that a large number of reaction rate constants are not available in the literature and difficult to be experimentally determined. Our extension saves the virtues of both BNs and rule-based kinetic modeling, while advancing the specification power to multicellular and multiscale models. We employ stochastic simulation NFsim \cite{sneddon2008nfsim} and statistical model checking (StatMC) \cite{jha2009bayesian} to analyze the system’s properties. The formal analysis results show that our model reproduces existing experimental findings with regard to the mutual promotion between pancreatic cancer and stellate cells. The model also provides insights into how treatments latching onto different targets could lead to distinct outcomes. Using the validated model, we predict novel (poly)pharmacological strategies for improving PC treatment. 

{\bf Related work.} Various mathematical formalisms have been used for the cancer microenvironment modeling (see a recent review \cite{altrock2015mathematics}). In particular, Gong \cite{gong2013analysis} built a qualitative model to analyze the intracellular signaling reactions in PCCs and PSCs. This model is discrete and focuses on cell proliferation, apoptosis, and angiogenesis pathways. While, our model is able to make quantitative predictions and also considers pathways regulating the autophagy of PCCs and the activation and migration of PSCs, as well as the interplay between PCCs and PSCs. In terms of the modeling language, the ML-Rules proposed by Maus et al. \cite{maus2011rule} extended BioNetGen to support hierarchical modeling. It allows users to describe inter- and intra-cellular processes at the cellular level. ML-Rules uses continuous rate equations to capture the dynamics of intracellular reactions, and thus requires all the rate constants to be known. Instead, our language models intracellular dynamics using BNs, which reduces the difficulty of estimating the values of hundreds of unknown parameters often involved in large models.


The paper is organized as follows. In Section 2, we present the multicellular model for the PC microenvironment. We then introduce our rule-based modeling formalism extended from the BioNetGen language in Section 3. In Section 4, we briefly introduce StatMC that is used to carry out formal analysis of the model. The analysis results are given and discussed in Section 5. Section 6 concludes the paper. 

\vspace{-.4cm}
\section{Reaction Network of Pancreatic Cancer Microenvironment}
\begin{figure}
\centering
\includegraphics[width=\linewidth]{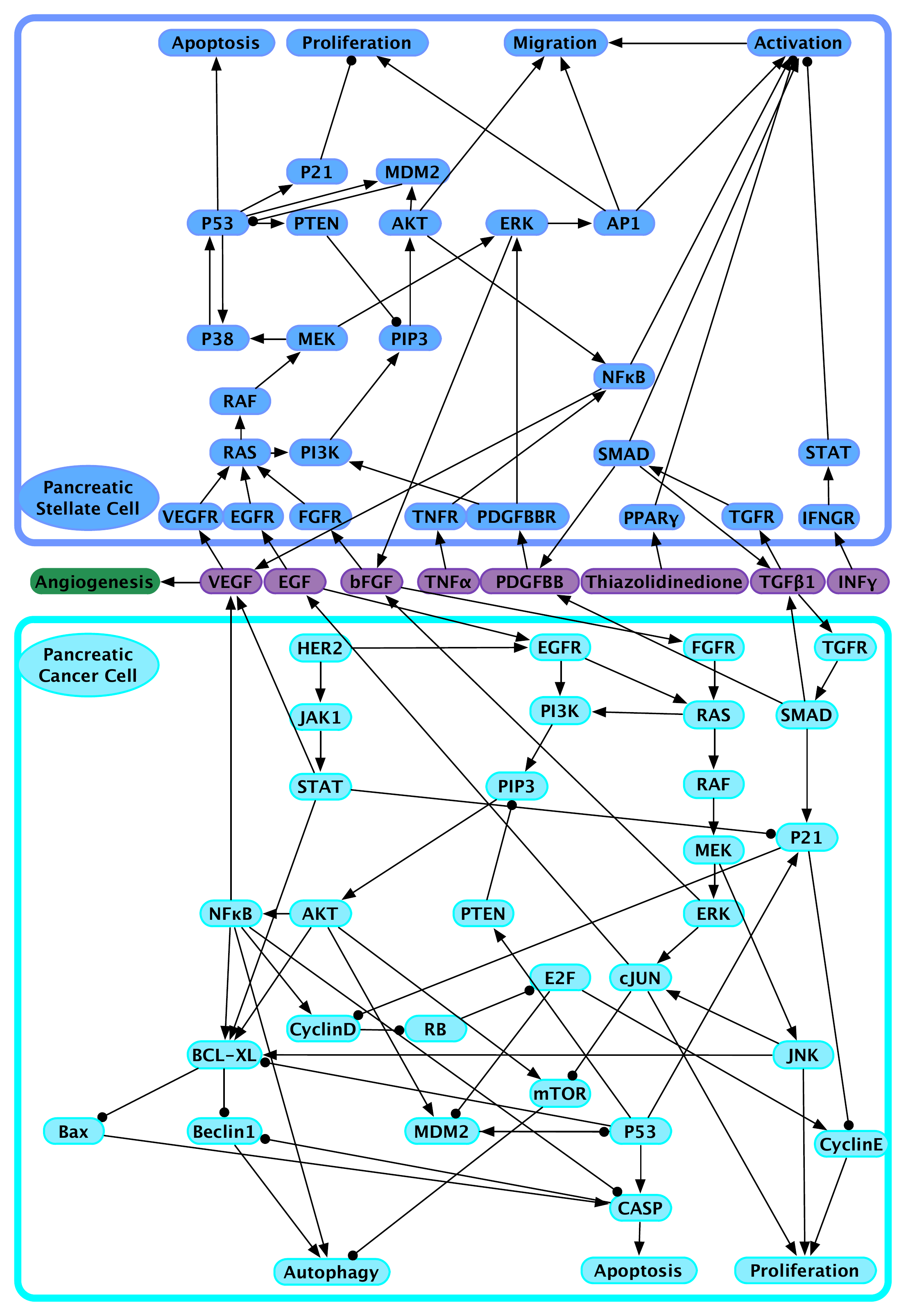}
\caption{The pancreatic cancer microenvironment model}
\label{fig:model}
\end{figure}

\vspace{-.3cm}
We construct a multicellular model for pancreatic cancer microenvironment based on a comprehensive literature search. The reaction network of the model is summarized in Figure \ref{fig:model}. It consists of three parts that are colored with green, blue, and purple respectively: (i) the intracellular signaling network of PCCs, (ii) the intracellular signaling network of PSCs, and (iii) the signaling molecules (such as growth factors and cytokines) in the extracellular space of the microenvironment, which are ligands of the receptors expressed in PCCs and PSCs. Note that $\to$ denotes activation/promotion/up-regulation, and $\multimapdot$ represents inhibition/ suppression/down-regulation. 


\vspace{-.5cm}
\subsection{The intracellular signaling network of PCCs}
\vspace{-.1cm}
\noindent\textit{\textbf{Pathways regulating proliferation}}

{\bf KRas mutation enhances proliferation} \cite{bardeesy2002pancreatic}. Mutations of the KRas oncogene occur in the precancerous stages with a mutational frequency over 90\%. It can lead to the continuous activation of the RAS protein, which then constantly triggers the RAF$\to$MEK cascade, and promotes PCCs' proliferation through the activation of ERK and JNK.

{\bf EGF activates and enhances proliferation} \cite{murphy2001pancreatic}. Epidermal growth factor (EGF) and its corresponding receptor (EGFR) are expressed in $\sim$95\% of PCs. EGF promotes proliferation through the RAS$\to$RAF$\to$MEK$\to$JNK cascade. It can also trigger the RAS$\to$RAF$\to$MEK$\to$ERK$\to$cJUN cascade to secrete EGF molecules, which can then quickly bind to overexpressed EGFR again to promote PCC proliferation, which is believed to confer the devastating nature on PCs.

{\bf HER2/neu mutation also intensifies proliferation} \cite{bardeesy2002pancreatic}. HER2/neu is another oncogene frequently mutated in the initial PC formation. Mutant HER2 can bind to EGFR to form a heterodimer, which can activate the downstream signaling pathways of EGFR. 

{\bf bFGF promotes proliferation} \cite{bensaid1992basic}. As a mitogenic polypeptide, bFGF can promote proliferation through both RAF$\to$MEK$\to$ERK and RAF$\to$MEK$\to$JNK cascades. In addition, bFGF molecules are released through RAF$\to$MEK$\to$ERK pathway to trigger another autocrine signaling pathway in the PC development.

\noindent\textit{\textbf{Pathways regulating apoptosis}}
 
Apoptosis is the most common mode of programmed cell death. It is executed by caspase proteases that are activated by death receptors or mitochondrial pathways.

{\bf TGF$\beta$1 initiates apoptosis} \cite{siegel2003cytostatic}. In PCCs, transforming growth factor $\beta$ 1 (TGF$\beta$1) binds to and activates its receptor (TGFR), which in turn activates receptor-regulated SAMDs that hetero-oligomerize with the common SAMD3 and SAMD4. After translocating to the nucleus, the complex initiates apoptosis in the early stage of the PC development.

{\bf Mutated oncogenes inhibit apoptosis}. Mutated KRas and HER2/neu can inhibit apoptosis by downregulating caspases (CASP) through PI3K$\to$AKT$\to$NF$\kappa$B cascade and by inhibiting Bax (and indirectly CASP) via PI3K$\to$PIP3$\to$AKT$\to$$\cdots$$\to$BCL-XL pathways.

\noindent\textit{\textbf{Pathways regulating autophagy}}

Autophagy is a catabolic process involving the degradation of a cell's own components through the lysosomal machinery. This pro-survival process enables a starving cell to reallocate nutrients from unnecessary processes to essential processes. Recent studies indicate that autophagy is important in the regulation of cancer development and progression and also affects the response of cancer cells to anticancer therapy \cite{kondo2005role,hippert2006autophagy}.

{\bf mTOR regulates autophagy} \cite{muilenburg2014role}. 
The mammalian target of rapamycin (mTOR) is a critical regulator of autophagy. In PCCs, the upstream pathway PI3K$\to$PIP3$\to$AKT activates mTOR and inhibits autophagy. The MEK$\to$ERK cascade downregulates mTOR via cJUN and enhances autophagy. 

{\bf Overexpression of anti-apoptotic factors promotes autophagy} \cite{marino2014self}. Apoptosis and autophagy can mutually inhibit each other due to their crosstalks. In the initial stage of PC, the upregulation of apoptosis leads to the inhibition of autophagy. Along with the progression of cancer, when apoptosis is suppressed by the highly expressed anti-apoptotic factors (e.g. NF$\kappa$B and Beclin1), autophagy gradually takes the dominant role and promotes PCC survival.

\vspace{-.4cm}
\subsection{Intracellular signaling network of PSCs}

\vspace{-.1cm}
\noindent\textit{\textbf{Pathways regulating activation}}

PCCs can activate the surrounding inactive PSCs by cancer-cell-induced release of mitogenic and fibrogenic factors, such as PDGFBB, TGF$\beta$1, and TNF$\alpha$. As a major growth factor regulating the cell functions of PSCs, {\bf PDGFBB activates PSCs}  \cite{haber1999activation} through the downstream ERK$\to$AP1 signaling pathway. The {\bf activation of PSCs is also mediated by TGF$\beta$1} \cite{haber1999activation} via TGFR$\to$SAMD pathway. The autocrine signaling of TGF$\beta$1 maintains the sustained activation of PSCs. Furthermore, the cytokine {\bf TNF$\alpha$ is also involved in activating PSCs} \cite{masamune2005ellagic} through binding to TNFR, which indirectly activates NF$\kappa$B.

\noindent\textit{\textbf{Pathways regulating migration}}

Migration is another characteristic cell function of PSCs. Activated PSCs move towards PCCs, and form a cocoon around tumor cells, which could protect the tumor from therapies' attacks \cite{feig2012pancreas, apte2004desmoplastic}.

{\bf Growth factors promote migration}. Growth factors existing in the microenvironment, including EGF, bFGF, and VEGF, can bind to their receptors on PSCs and activate the migration through the MAPK pathway. 

{\bf PDGFBB contributes to the migration} \cite{phillips2003cell}. PDGFBB regulates the migration of PSCs mainly through two downstream pathways: (i) the PI3K$\to$PIP3$\to$AKT pathway, which mediates PDGF-induced PSCs' migration, but not proliferation, and (ii) the ERK$\to$AP1 pathway that regulates activation, migration, and proliferation of PSCs. 

\noindent\textit{\textbf{Pathways regulating proliferation}}

{\bf Growth factors activate proliferation}. In PSCs, as key downstream components for several signaling pathways initiated by distinct growth factors, such as EGF and bFGF, the ERK$\to$AP1 cascade activates the proliferation of PSCs. Compared to inactive PSCs, active ones proliferate more rapidly.

{\bf Tumor suppressers repress proliferation}. Similar to PCCs, P53, P21, and PTEN act as suppressers for PSCs' proliferation. 

\noindent\textit{\textbf{Pathways regulating apoptosis}}

{\bf P53 upregulates modulator of apoptosis} \cite{jaster2004molecular}. The apoptosis of PSCs can be initiated by P53. The expression of p53 is regulated by the MAPK pathway.

\vspace{-.4cm}
\subsection{Interactions between PCCs and PSCs}

\vspace{-.1cm}
The mechanism underlying the interplay between PCCs and PSCs is complex. In a healthy pancreas, PSCs exist quiescently in the periacinar, perivascular, and periductal space. However, in the diseased state, PSCs will be activated by growth factors, cytokines, and oxidant stress secreted or induced by PCCs, including EGF, bFGF, VEGF, TGF$\beta$1, PDGF, sonic hedgehog, galectin 3, endothelin 1 and serine protease inhibitor nexin 2 \cite{duner2011pancreatic}. Activated PSCs will then transform from the quiescent state to the myofibroblast phenotype. This results in their losinlipid droplets, actively proliferating, migrating, producing large amounts of extracellular matrix, and expressing cytokines, chemokines, and cell adhesion molecules. In return, the activated PSCs promote the growth of PCCs by secreting various factors, including stromal-derived factor 1, FGF, secreted protein acidic and rich in cysteine, matrix metalloproteinases, small leucine-rich proteoglycans, periostin and collagen type I that mediate effects on tumor growth, invasion, metastasis and resistance to chemotherapy \cite{duner2011pancreatic}. Among them, EGF, bFGF, VEGF, TGF$\beta$1, and PDGFBB are essential mediators of the interplay between PCCs and PSCs that have been considered in our model. 

{\bf Autocrine and paracrine involving EGF/bFGF} \cite{mahadevan2007tumor}. EGF and bFGF can be secreted by both PCCs and PSCs. In turn, they will bind to EGFR and FGFR respectively on both PCCs and PSCs to activate their proliferation and further secretion of EGF and FGF.

{\bf Interplay through VEGF} \cite{vonlaufen2008pancreatic}. As a proangiogenic factor, VEGF is found to be of great importance in the activation of PSCs and angiogenesis during the progression of PCs. VEGF, secreted by PCCs, can bind with VEGFR on PSCs to activate the PI3K pathway. It further promotes the migration of PSCs through PIP3$\to$AKT, and suppresses the transcription activity of P53 via MDM2.

{\bf Autocrine and paracrine involving TGF$\beta$1} \cite{mahadevan2007tumor}. PSCs by themselves are capable of synthesizing TGF$\beta$1, suggesting the existence of an autocrine loop that may contribute to the perpetuation of PSC activation after an initial exogenous signal, thereby promoting the development of pancreatic fibrosis.

{\bf Interplay through PDGFBB} \cite{duner2011pancreatic}. PDGFBB exists in the secretion of PCCs, whose production is regulated by TGF$\beta$1 signaling pathway. PDGFBB can activate PSCs and initiate migration and proliferation as well.

\vspace{-.4cm}
\section{The Modeling Language}
\vspace{-.2cm}
Rule-based modeling languages are often used to specify protein-to-protein reactions within cells and to capture the evolution of protein concentrations. BioNetGen language is a representative rule-based modeling formalism \cite{faeder2009rule}, which consists of three components: basic building blocks, patterns, and rules. In our setting, in order to simultaneously simulate the dynamics of multiple cells, interactions among cells, and intracellular reactions, we advance the specifying power of BioNetGen by redefining basic building blocks and introducing new types of rules for cellular behaviors as follows.

{\bf Basic building blocks}. The basic building blocks in BioNetGen are molecules that may be assembled into complexes through bonds linking components of different molecules. To handle multiscale dynamics (i.e. cellular and molecular levels), we allow the fundamental blocks to be also cells or extracellular molecules. Specifically, a cell is treated as a fundamental block with subunits corresponding to the components of its intracellular signaling network. Furthermore, extracellular molecules (e.g. EGF) are treated as fundamental blocks without subunits. 

Since we use BNs to model intracellular signaling networks, each subunit of a cell takes binary values (it is straightforward to extend BNs to discrete models). The Boolean values - ``True (T)'' and ``False (F)'' - can have different biological meanings for distinct types of components within the cell. For example, for a subunit representing cellular process (e.g. apoptosis), ``T'' means the cellular process is triggered, and ``F'' means it is not triggered. For a receptor, ``T'' means the receptor is bound, and ``F'' means it is free. For a protein, ``T'' indicates this protein has a high concentration, and ``F'' indicates that its concentration level is below the value to regulate the downstream targets. 

{\bf Patterns}. As defined in BioNetGen, patterns are used to identify a set of species that share features. For instance,  the pattern $C(c_1)$ matches both $C(c_1, c_2 \sim T)$ and $C(c_1, c_2 \sim F)$. Using patterns offers a rich yet concise description in specifying components.

{\bf Rules}. In BioNetGen, three types of rules are used to specified: binding/unbinding, phosphorylation, and dephosphorylation. Here we introduce nine rules in order to describe the cellular processes in our model and the potential therapeutic interventions. For each type of rules, we present its formal syntax followed by examples that demonstrate how it is used in our model.

\noindent{\it \bf Rule 1: Ligand-receptor binding}
\[<Lig> + <Cell>(<Rec> \sim F)  \to <Cell>(<Rec> \sim T) \;\;\;\;  <binding\_rate>\]
\textit{Remark}: 
On the left-hand side, the ``F'' value of a receptor $<Rec>$ indicates that the receptor is free. When a ligand $<Lig>$ binds to it, the reduction of number of extracellular ligand is represented by its elimination. In the meanwhile, ``$<Rec> \sim T$'', on the right-hand side, indicates that the receptor is not free any more. Note that, the multiple receptors on the surface of a cell can be modeled by setting a relatively high rate on the following downstream regulating rules, which indicates the rapid ``releasing'' of bound receptors. An example in our microenvironment model is the binding between EGF and EGFR for PCCs: ``$EGF+PCC(EGFR \sim F) \to PCC(EGFR \sim T) \;\; 1$''.

\noindent{\it \bf Rule 2: Mutated receptors form a heterodimer}
\[\hspace{-5cm} <Cell>(<Rec_1> \sim F, <Rec_2> \sim F) \to \]
\vspace{-.7cm}
\[\hspace{2cm} <Cell>(<Rec_1> \sim T, <Rec_2> \sim T)  \;\;\;\; <mutated\_binding\_rate>\]
\textit{Remark}: 
Unbound receptors can bind together and form a heterodimer. For example, in our model, the mutated HER2 can activate downstream pathways of EGFR by binding with it and forming a heterodimer: ``'$PCC(EGFR \sim F,HER2 \sim F) \to PCC(EGFR \sim T,HER2 \sim T) \;\; 10$''.

\noindent{\it \bf Rule 3: Downstream signaling transduction}

\noindent Rule 3.1 (Single parent) upregulation (activation, phosphorylation, etc.)
\[<Cell>(<Act> \sim T, <Tar> \sim F) \to <Cell>(<Act> \sim T, <Tar> \sim T)  \;\;  <trate>\]
\noindent Rule 3.2 (Single parent) downregulation (inhibition, dephosphorylation, etc.)
\[<Cell>(<Inh> \sim T, <Tar> \sim T) \to <Cell>(<Inh> \sim T, <Tar> \sim F)   \;\;  <trate>\]
\noindent Rule 3.3 (Multiple parents) Downstream regulation
\[\hspace{-3.7cm} <Cell>(<Inh> \sim F, <Act> \sim T, <Tar> \sim F) \to \]
\vspace{-.7cm}
\[\hspace{3cm} <Cell>(<Inh> \sim F, <Act> \sim T, <Tar> \sim T)  \;\; <trate>\]
\[<Cell>(<Inh> \sim T, <Tar> \sim T) \to <Cell>(<Inh> \sim T, <Tar> \sim F)  \;\; <trate>\]
\textit{Remark}: Instead of using kinetic rules (such as in ML-Rules), our language use logical rules of BNs to describe intracellular signal cascades. Downsteam signal transduction rules are used to describe the logical updating functions for all intracellular molecules constructing the signaling cascades. For instance, Rule 3.3 presents the updating function  $<Tar>^{(t+1)} = \neg <Inh>^{(t)} \times  {(<Act>^{(t)}} + <Tar>^{(t)})$, where ``$<Inh>$'' is the inhibitor, and ``${<Act>}$'' is the activator. In this manner, concise rules can be devised to handle complex cases, where there exists multiple regulatory parents. Note that our model follows the biological assumption that inhibitors hold higher priorities than activators with respect to their impacts on the target. ``+'' and ``$\times$'' in logical functions represent logical ``OR'' and ``AND'' respectively. An example in our model is that, in PCCs, $STAT$ can be activated by $JAK1$: ``$PCC(JAK1\sim T,STAT\sim T) \to PCC(JAK1\sim F,STAT\sim T) \;\; 0.012$'' and ``$PCC(JAK1\sim T,STAT\sim F) \to PCC(JAK1\sim F,STAT\sim T) \;\; 0.012$''.

\noindent{\it \bf Rule 4: Cellular processes}

\noindent Rule 4.1 Proliferation
\[<Cell>(Pro \sim T) \to <Cell>(Pro \sim F)+ <Cell>(Pro \sim F, \cdots)  \;\;\;\; <pro\_rate>\]
\textit{Remark}: When a cell proliferates, we keep the current values of subunits for the cell that initiates the proliferation, and assume the new cell to have the default values of subunits. The ``$\cdots$'' in the rule denotes the remaining subunits with their default values. 

\noindent Rule 4.2 Apoptosis
\[<Cell>(Apo \sim T) \to Null()  \;\;\;\;  <apop\_rate>\]
\textit{Remark}: A type ``Null()'' is declared to represent dead cells or degraded molecules. In our model, the apoptosis of PSCs is described as ``$PSC(Apo \sim T) \to Null() \;\; 5e-4$''.

\noindent Rule 4.3 Autophagy
\[<Cell>(Aut \sim T) \to <Mol> + \cdots   \;\;\;\;  <auto\_rate>\]
\textit{Remark}: The molecules on the right-hand side of this type of rules will be released into the microenvironment due to autophagy. They are the existing molecules expressed inside this cell when autophagy is triggered.

\noindent{\it \bf Rule 5: Secretion}
\[<Cell>(<secMol> \sim T) \to <Cell>(<secMol> \sim F)+ <Mol> \;\;\;\; <sec\_rate>\]
\textit{Remark}: When the secretion of ``$<Mol>$'' has been triggered, its amount in the microenvironment will be added by 1. Note that, we can differentiate the endogenous and exogenous molecules by labeling the secreted ``$<Mol>$'' with the cell name. In our model, we have ``$PCC(secEGF \sim T) \to PCC(secEGF \sim F)+EGF \;\; 2.7e-4$''.

\noindent{\it \bf Rule 6: Mutation}
\[<Cell>(<Mol> \sim <unmutated>) \to <Cell>(<Mol> \sim <mutated>) \;\;   <mrate>\] 
\textit{Remark}: For mutant proteins that are constitutively active, we set a very high value to the mutation rate ``mrate''. In this way, we can almost keep the value of the mutated molecule as what it should be. For example, in our model, the mutation of oncoprotein $Ras$ in PCCs is captured by ``$PCC(RAS \sim F) \to PCC(RAS \sim T) \;\; 10000$''.

\noindent{\it \bf Rule 7: Constantly over-expressed extracellular molecules}
\[CancerEvn \to CancerEvn + <Mol>  \;\;\;\;  <sec\_rate>\]
\textit{Remark}: We use this type of rules to mimic the situation that the concentration of an over-expressed extracellular molecule stays in a high level constantly. 

\noindent{\it \bf Rule 8: Degradation of extracellular molecules}
\[<Mol> \to Null()  \;\;\;\; <deg\_rate>\]
\textit{Remark}: Here, ``Null()'' is used to represent dead cells or degraded molecules. For instance, $bFGF$ in the microenvironment will be degraded via ``$bFGF \to Null() \;\; 0.05$''.

\noindent{\it \bf Rule 9: Therapeutic intervention}
\[<Cell>(<Mol> \sim <untreated>) \to <Cell>(<Mol> \sim <treated>) <treat\_rate>\] 
\textit{Remark}: Given a validated model, intervention rules allow us to evaluate the effectiveness of a therapy targeting at certain molecule(s). Also, the well-tuned value of the intervention rate can, more or less, give indications when deciding the dose of medicine used in this therapy, based on the Law of Mass Action.

Our extension allows the BioNetGen language to be able to model not only the signaling network within a single cell, but also interactions among multiple cells. It also allows one to simulate the dynamics of cell populations, which is crucial to cancer study. Moreover, describing the intracellular dynamics using the style of BNs improves the scalability of our method by overcoming the difficulty of obtaining values of a large amount of model parameters from wet laboratory, which is a common bottleneck of conventional rule-based languages and ML-Rules. Note that, similar to other rule-based languages, our extended one allows different methods for model analysis, since more than one semantics can be defined for the same syntax.

\vspace{-.4cm}
\section{Statistical Model Checking}
\vspace{-.3cm}
Simulation can recapitulate a number of experimental observations and provide new insights into the system. However, it is not easy to manually analyze a significant amount of simulation trajectories, especially when there is a large set of system properties to be tested. Thus, for our model, we employ statistical model checking (StatMC), which is a fully automated formal analysis technique. In this section, we provide an intuitive and brief description of StatMC. The interested reader can find more details in \cite{jha2009bayesian}.

Given a system property expressed as a Bounded Linear Temporal Logic (BLTL) \cite{jha2009bayesian} formula and the set of simulation trajectories generated by applying the NFsim stochastic simulation to our rule-based model, StatMC estimates the probability of the model satisfying the property. (See Appendix \ref{apndx:bltl} for a brief introduction of BLTL.) In detail, since the underlying semantic model of the stochastic simulation method NFsim that we used for our model is essentially a discrete-time Markov chain, we need to verify stochastic models. StatMC treats the verification problem for stochastic models as a statistical inference problem, using randomized sampling to generate traces (or simulation trajectories) from the system model, and then performing model checking and statistical analysis on those traces. For a (closed) stochastic model and a BLTL property $\psi$, the probability $p$ that the model satisfies $\psi$ is well defined (but unknown in general). For a fixed $0< \theta <1$, we ask whether $p \le \theta$, or what the value of $p$ is. In StatMC, the first question is solved via hypothesis testing methods, while the second via estimation techniques. Intuitively, hypothesis tests are probabilistic decision procedures, i.e., algorithms with a yes/no reply, and which may give wrong answers. Estimation techniques instead compute (probabilistic) approximations of the unknown probability $p$. The main assumption of StatMC is that, given a BLTL property $\psi$, the behavior of a (closed) stochastic model can be described by a Bernoulli random variable of parameter $p$, where $p$ is the probability that the system satisfies $\psi$. It is known that discrete-time Markov chains satisfy this requirement \cite{vardi1985automatic}. Therefore StatMC can be applied to our setting. More specifically, given $\sigma$ is a system execution and $\psi$ a BLTL formula, we have that $Prob\{ \sigma | \sigma \models \psi \} = p$, and the Bernoulli random variable mentioned above is the following function $M$ defined as follows: $M(\sigma) = 1$ if $\sigma \models \psi$, or $M(\sigma) = 0$ otherwise. Therefore, $M$ will be 1 with probability $p$ and 0 with probability $1-p$. In general, StatMC works by first obtaining samples of $M$, and then by applying statistical techniques to such samples to solve the verification problem. 


\vspace{-.3cm}
\section{Results and Discussion}
\vspace{-.2cm}
In this section, we present and discuss formal analysis results for our pancreatic cancer microenvironment model. The model file is available at \url{http://www.cs.cmu.edu/~qinsiw/mpc_model.bngl}. All the experiments reported below were conducted on a machine with a 1.7 GHz Intel Core i7 processor and 8GBRAM, running on Ubuntu 14.04.1 LTS. In our experiments, we use Bayesian sequential estimation with 0.01 as the estimation error bound, coverage probability 0.99, and a uniform prior ($\alpha$ = $\beta$ = 1). The time bounds and thresholds given in following properties are determined by considering the model's simulation results. The parameters in our model include initial state (e.g. abundance of extracellular molecules) and reaction rate constants. The initial state was provided by biologists based on wet-lab measurements. The rate constants were estimated based on the general ones in the textbook \cite{alon2006introduction}. The results in scenario I \& II demonstrate that using these parameters the model is able to reproduce key observations reported in the literature. We also performed a sensitivity analysis and the results show that the system behavior is robust to most of the parameters (the two sensitive parameters have been labeled in our model file).

\noindent{\bf Scenario I: mutated PCCs with no treatments}


In scenario I, we validate our model by studying the role of PSCs in the PC development.

\noindent {\bf Property 1}: This property aims to estimate the probability that the population of PCCs will eventually reach and maintain in a high level.
\[Prob_{= ?} \; \{(PCCtot = 10) \wedge F^{1200} \; G^{100} \; (PCCtot > 200)\}\]
First, we take a look at the impact from the presence of PSCs on the dynamics of PCC population. As shown in Table \ref{table:res}, with PSCs, the probability of the number of PCCs reaching and keeping in a high level ($Pr=0.9961$) is much higher than the one when PSCs are absent ($Pr=0.405$). This indicates that PSCs promote PCCs proliferation during the progression of PC. This is consistent with experimental findings \cite{apte2004desmoplastic, duner2011pancreatic, vonlaufen2008pancreatic}.

\vspace{-.5cm}
\begin{table}[!ht]
\captionsetup{font=scriptsize}
\centering
\begin{tabular}{|c|c|c|c|c|p{4.5cm} < {\centering}|}
\hline
Property & Estimated Prob & \# Succ & \# Sample & Time (s) & Note\\ \hline \hline
\multicolumn{6}{|c|}{Scenario I: mutated PCCs with no treatments} \\ \hline
1 & 0.4053 & 10585 & 26112 & 208.91 & w.o. PSCs \\ \hline
{} & 0.9961 & 256 & 256 & 1.83  & w. PSCs\\ \hline
2 & 0.1191 & 830 & 6976 & 49.69 & w.o. PCCs \\ \hline
{} & 0.9961 & 256 & 256 & 1.75 & w. PCCs  \\ \hline
3 & 0.9961 & 256 & 256 & 5.21 & {-}  \\ \hline
4 & 0.9961 & 256 & 256 & 4.38 & {-} \\ \hline \hline
\multicolumn{6}{|c|}{Scenario II: mutated PCCs with different exsiting treatments} \\ \hline
5 & 0.0004 & 0 & 2304 & 17.13 & cetuximab and erlotinib\\ \hline
{} & 0.0012 & 10 & 9152 & 68.67 & gemcitabine \\ \hline
{} & 0.7810 & 8873 & 11360 & 114.25 & nab-paclitaxel \\ \hline
{} & 0.8004 & 7753 & 9686 & 73.83 & ruxolitinib \\ \hline \hline
\multicolumn{6}{|c|}{Scenario III: mutated PCCs with blocking out on possible target(s)} \\ \hline
6 & 0.0792 & 38363 & 484128 & 3727.99 & w.o. inhibiting ERK in PSCs\\ \hline
{} & 0.9822 & 2201 & 2240 & 17.37 & w. inhibiting ERK in PSCs \\ \hline
7 & 0.1979 & 3409 & 17232 & 136.39 & w.o. inhibiting ERK in PSCs \\ \hline
{} & 0.9961 & 256 & 256 & 2.01 & w. inhibiting ERK in PSCs \\ \hline
8 & 0.2029 & 2181 & 10752 & 92.57 & w.o. inhibiting MDM2 in PSCs \\ \hline
{} & 0.9961 & 256 & 256 & 2.18 & w. inhibiting MDM2 in PSCs \\ \hline
9 & 0.0004 & 0 & 2304 & 15.77 & w.o. inhibiting RAS in PCCs and ERK in PSCs  \\ \hline
{} & 0.9961 & 256 & 256 & 3.15 & w. inhibiting RAS in PCCs and ERK in PSCs  \\ \hline
10 & 0.9797 & 1349 & 1376 & 11.98 & w.o. inhibiting STAT in PCCs and NF$\kappa$B in PSCs \\ \hline
{} & 0.1631 & 1476 & 9056 & 81.61 & w. inhibiting STAT in PCCs and NF$\kappa$B in PSCs  \\ \hline
\end{tabular}
\caption{Statistical model checking results for properties under different scenarios}
\vspace{-.7cm}
\label{table:res}
\end{table}

\noindent {\bf Property 2}: This property aims to estimate the probability that the number of migrated PSCs will eventually reach and maintain in a high amount.
\[Prob_{= ?} \; \{(MigPSC = 0)  \wedge F^{1200} \; G^{100} \; (MigPSC > 40)\}\]
We then study the impacts from PCCs on PSCs. As shown in Table \ref{table:res}, without PCCs, it is quite unlikely (($Pr = 0.1191$) for quiescent PSCs to be activated. While, when PCCs exist, the chance of PSCs becoming active (($Pr=0.9961$) approaches to 1. This confirms the observation \cite{haber1999activation} that, during the development of PC, PSCs will be activated by growth factors, cytokines, and oxidant stress secreted or induced by PCCs.

\noindent {\bf Property 3}: This property aims to estimate the probability that the number of PCCs entering the apoptosis phase will be larger than the number of PCCs starting the autophagy process and this situation will be reversed eventually.
\[Prob_{= ?} \; \{F^{400} \; (G^{300} \; (ApoPCC > 50 \wedge AutoPCC < 50) \vspace{-.3cm}\]

\hspace{2.5cm} $\wedge F^{700} \; G^{300} \; (ApoPCC < 50 \wedge AutoPCC > 50))\}$

\vspace{.15cm}
\noindent We are also interested in the mutually exclusive relationship between apoptosis and autophagy for PCCs reported in \cite{hippert2006autophagy, marino2014self}. In detail, as PC progresses, apoptosis firstly overwhelms autophagy, and then autophagy takes the leading place after a certain amount of time. This situation is described as property 3 and its estimated probability is close to 1 (see Table \ref{table:res}).

\noindent {\bf Property 4}: This property aims to estimate the probability that, it is always the case that, once the population of activated PSCs reaches a high level, the number of migrated PSCs will also increase.
\[Prob_{= ?} \; \{G^{1600} \; (ActPSC > 10 \to F^{100} \; (MigPSC > 10))\}\]
One reason why PC is hard to be cured is that activated PSCs will move towards mutated PCCs, and form a cocoon for the tumor cells, which can protect tumor from attacks caused by therapies \cite{feig2012pancreas, apte2004desmoplastic}. We investigate this by checking property 4, and obtain an estimated probability approaching to 1 (see Table \ref{table:res}).

\noindent{\bf Scenario II: mutated PCCs with different existing treatments}

\noindent {\bf Property 5}: This property aims to estimate the probability that the population of PCCs will eventually drop to and maintain in a low amount.
\[Prob_{= ?} \; \{(PCCtot = 10) \wedge F^{1200} \; G^{400} \; (PCCtot < 100)\}\]
Property 5 means that, after some time, the population of PCCs can be maintained in a comparatively low amount, implying that PC is under control. We now consider 
5 different drugs that are widely used in PC treatments - cetuximab, erlotinib, 
gemcitabine, nab-paclitaxel, and ruxolitinib, and estimate the probabilities for them to satisfy property 5. As shown in Table \ref{table:res}, monoclonal antibody targeting EGFR (cetuximab), as well as direct inhibition of EGFR (erlotinib) broadly do not provide a survival benefit in PCs. 
Inhibition of MAPK pathway (gemcitabine) has also not been promising. 
These results are consistent with clinical feedbacks from patients \cite{edson}. While, strategies aiming at depleting the PSCs in PCs (i.e. nab-paclitaxel) can be successful (with an estimated probability 0.7810). Also, inhibition of Jak/Stat can be very promising (with an estimated probability 0.8004). These results are supported by \cite{von2013increased} and \cite{hurwitz2014randomized}, respectively.

\noindent{\bf Scenario III: mutated PCCs with blocking out on possible target(s)}


Scenario I and II have demonstrated the descriptive and predictive power of our model.  In scenario III, we use the validated model to identify new therapeutic strategies targeting molecules in PSCs. Here we report 4 potential target(s) of interest from our screening.
 
\noindent {\bf Property 6}: This property aims to estimate the probability that the number of PSCs will eventually drop to and maintain in a low level.
\[Prob_{= ?} \; \{(PSCtot = 5) \wedge F^{1200} \; G^{400} \; (PSCtot < 30)\}\]
\noindent {\bf Property 7}: This property aims to estimate the probability that the population of migrated PSCs will eventually stay in a low amount.
\[Prob_{= ?} \; \{(MigPSC = 0)  \wedge F^{1200} \; G^{100} \; (MigPSC < 30)\}\]
The verification results of these two properties (Table \ref{table:res}) suggest that inhibiting ERK in PSCs not only lowers the population of PSCs, but also inhibits PSC migration. The former function can reduce the assistance from PSCs in the progression of PCs indirectly. The later one can prevent PSCs from moving towards PCCs and forming a cocoon to protect PCCs against cancer treatments.

\noindent {\bf Property 8}: This property aims to estimate the probability that the number of PSCs entering the proliferation phase will eventually be less than the number of PSCs starting the apoptosis programme and this situation will maintain.
\[Prob_{= ?} \; \{F^{1200} \; G^{400} \; ((PSCPro - PSCApop) < 0)\}\]
The increased probability (from 0.2029 to 0.9961 as shown in Table \ref{table:res}) indicates that inhibiting MDM2 in PSCs may reduce the number of PSCs by inhibiting PSCs' proliferation and/or promoting their apoptosis. Similar to the former role of inhibiting ERK in PSCs, it can help to treat PCs by alleviating the burden caused by PSCs.

\noindent {\bf Property 9}: This property aims to estimate the probability that the number of bFGF will eventually stay in such a low level.
\[Prob_{= ?} \; \{F^{1200} \; G^{400} \; (bFGF < 100)\}\]
As mentioned in property 5, 6, and 7, inhibiting RAS in PCCs can lower the number of PCCs, and downregulating ERK in PSCs can inhibit their proliferation and migration. Besides these, we find that, when inhibiting RAS in PCCs and ERK in PSCs simultaneously, the concentration of bFGF in the microenvironment drops (see Table \ref{table:res}). As bFGF is a key molecule that induces proliferation of both cell types, targeting RAS in PCCs and ERK in PSCs at the same time could synergistically improve  PC treatment.

\noindent {\bf Property 10}: This property aims to estimate the probability that the concentration of VEGF will eventually reach and keep in a high level.
\[Prob_{= ?} \; \{F^{400} \; G^{100} \; (VEGF > 200)\}\]
Furthermore, inhibiting STAT in PCCs and NF$\kappa$B in PSCs simultaneously postpones and lowers the secretion of VEGF (see Table \ref{table:res}). VEGF plays an important role in the angiogenesis and metastasis of pancreatic tumors. Thus, the combinatory inhibition of STAT in PCCs and NF$\kappa$B in PSCs may be another potential strategy for PC therapies.

\vspace{-.4cm}
\section{Conclusion}
\vspace{-.3cm}
We present a multicellular and multiscale model of the PC microenvironment. The model is formally described using the extended BioNetGen language, which enables us to capture the dynamics of multiscale biological systems using a combination of continuous and discrete rules. We carry out stochastic simulation and StatMC to analyze system behaviors under diffident conditions. Our verification results confirm the experimental findings with regard to the mutual promotion between PCCs and PSCs. We also gain insights on how existing treatments latching onto different targets can lead to distinct outcomes. These results demonstrate that our model could be used as a prognostic platform to identify new drug targets. We then identify four potentially (poly)pharmacological strategies for depleting PSCs and inhibiting the PC development. We plan to test our predictions empirically in future. Another interesting direction is to extend the model by taking account of cancer-associated macrophages in the PC microenvironment.

\vspace{-.2cm}
\bibliography{ref}{}
\bibliographystyle{abbrv}

\newpage
\appendix
\section{Bounded Linear Temporal Logic}\label{apndx:bltl}

In this appendix, we will briefly review Bounded LTL that we use to encode system properties in the ``Results and Discussion'' section. 

Linear Temporal Logic (LTL) \cite{pnueli1977temporal}, as a modal temporal logic with modalities referring to time, is widely used to formally encode formulae about the future of paths, such as a condition will eventually be true or a condition will be true until another fact becomes true. 
Bounded Linear Temporal Logic (BLTL) extends LTL with time bounds on temporal operators. 
For example, the following BLTL formula can be used to express the specification “it is not the case that within 5 seconds, variable $v_0$ will keep the value 1 and variable $v_1$ will keep the value 0 for 10 seconds”. 
\[ \neg F^5 G^{10} (v_0 =1 \wedge v_1 = 0) \]
where the $F^5$ operator encodes “future 5 seconds”, $G^{10}$ expresses “globally for 10 seconds”, and $v_0$ and $v_1$ are state variables of the model. 

The syntax of BLTL is given by:
\[\psi ::= x \sim v| \neg \psi | \psi_1 \vee \psi_2 | \psi_1 U^t \psi_2\]
where $x \in SV$ (the finite set of state variables), $\sim \in \{ <, \le, =, \ge, > \}$, $v \in \mathbb{Q}$, and $t \in \mathbb{Q}_{\ge 0}$. Note that the operators $\wedge$, $F^t$, and $G^t$ referenced above can be defined as follows:
$F^t \psi = True \; U^t \psi$, $G^t \psi = \neg F^t \neg \psi$, and $\psi_1 \wedge \psi_2 = \neg(\neg \psi_1 \vee \neg \psi_2)$

The semantics of BLTL is defined with respect to traces (or executions) of the model. For this work, a trace will be a simulation trajectory of our multiscale hybrid rule-based model. Formally, a trace is a sequence of time-stamped state transitions of the form $\sigma = (s_0,t_0), (s_1, t_1), \cdots$, indicating that the system moved to state $s_{i+1}$ after duration $t_i$ in state $s_i$. The fact that a trace $\sigma$ satisfies the BLTL property $\psi$ is denoted by $s \models \psi$. We denote the execution trace starting at state $i$ by $\sigma^i$. The value of the state variable $x$ in $\sigma$ at the state $i$ is denoted by $V (\sigma, i, x)$. The semantics of BLTL for a trace $\sigma^k$ starting at the kth state ($k \in N$) is defined as follows.
\begin{itemize}
\item $\sigma^k \models x \sim v$ if and only if $V (\sigma, i, x) \sim v$;
\item $\sigma^k \models \neg \psi$ if and only if $\sigma^k \models \psi$ does not hold;
\item $\sigma^k \models \psi_1 \vee \psi_2$ if and only if $\sigma^k \models \psi_1$ or $\sigma^k \models \psi_2$;
\item $\sigma^k \models \psi_1 U^t \psi_2$ if and only if there exists $i \in \mathbb{N}^+$ such that (a) $\sum_{j=k}^{k+i-1}t_j \le t $, (b) $\sigma^{k+i} \models \psi_2$, and (c) for each $0 \le j < i$, $\sigma^{k+j} \models \psi_1$.
\end{itemize}


\end{document}